\documentclass[11pt]{article}
  \usepackage{amssymb}
   \usepackage{graphicx}
\def\ba{\begin{eqnarray}}
\def\bam{\begin{array}}

\def\be{\begin{equation}}

\def\bi{\bibitem}

\def\bt {\beta}

\def\B {\overline}

\def\Br{\B r}

\def\ea{\end{eqnarray}} 
\def\ee{\end{equation}}

\def\fr{\frac}

\def\ha{\frac{1}{2}~}

\def\la {\lambda}
\def\lb{\label}

\def\nn{\nonumber}

\def\ts{\textstyle}

\def\1{{\it one}}

\def\2{{\ts{\ha}\!}}

\def\3 {\ts{\frac{1}{3}\!}}
\def\4{\ts{\fr{1}{4}\!}}

\begin{document}
\title{Variations on the theme of Michel H\'enon's Isochrone}
\author{ D.Lynden-Bell$^{1,2}$ 
%\thanks{email:dlb@ast.cam.ac.uk}
\\  {\it$^1$  Institute of Astronomy, The Observatories,
 Madingley Road, Cambridge CB3 0HA}\\
$^2$ Visiting {\it The Racah Institue of the Hebrew University of Jerusalem}}
\date{}      
\maketitle                                     % Activate to display a given date or no date
%\end{document}
{\bf Abstract} A variation of Newton's method of mapping Kepler's orbits into orbits in the
simple harmonic oscillator is shown to give H\'enon's Isochrone. The statistical mechanics of a micro-canonical ensemble of isochrone oscillators shows that the temperature reaches a maximum as a function of the energy and then declines to zero at the escape energy. In that declining region adding heat (energy) decreases the temperature, as occurs in star clusters. We then define the internal temperature of an ensemble of binary stars all at the same (negative) energy and show that they too get cooler when heated and hotter when cooled. When the internal  temperature of a binary is less that the temperature of the stars it interacts with, then on average heat will flow into it, making it less bound and of still lower temperature. Conversely hard binaries have higher internal temperatures than the local stars, so they lose energy and become hotter and yet more strongly bound, a process invoked by Michel H\'enon in his explanation of star-cluster evolution. Finally we give an isochronal variation of Newton's exactly soluble N-body problem.
%Introduction
\section{Introduction}
Michel H\'enon had a mind of wonderful clarity. In treating difficult and even intractable problems
he would pick out a simple special case which still contained the essential characteristics of the
general problem. His detailed studies of such special cases gave the insight that enabled him
to explain the essential nature of the solutions to the general problem. He was a wonderfully clear lecturer in both French and English greatly helped by his full command of both languages and excellent elocution.

	My work was strongly influenced by his in the years 1960-1980 when we both worked on
the stellar dynamics of galaxies and star clusters. On reading my paper on Violent Relaxation \cite{LB}
Michel noticed that I gave an incorrect characterisation of a general diffusive process. Iin the subsequent correspondence he was the first to give the correct criterion which Tremaine 
found independently many years later \cite{THL}. If we call the conserved thing that is diffusing 
mass then we can characterise different regions by their density. Now choose a density and include all mass in regions of higher density but exclude that in regions of lower density. Let the total included mass be m and the total volume of the included regions be V(m). After doing this for all possible choices of the density the function V(m) is found for any distribution.  His criterion for a diffusive process is that V(m) must increase, or at least not decrease, due to diffusion at every value of m. This criterion already shows that precision of thought that characterises his work. 

Isaac Newton in his Principia \cite{Ne} solved an N-body problem for all N, all initial conditions and all mass ratios, but to do this he took the linear force law between all masses $G'm_1m_2 {\bf (r_2-r_1)}$.
We shall return to this at the conclusion.
	Here I shall discuss some new results that relate the Isochrone both to Newton's Principia  \cite{Ne}, as discussed in Chandrasekhar's fine book on it \cite{Ch}, and to the statistical mechanics of binary stars which drive the evolution in H\'enon's \cite{He} model of star clusters.
	%Mapping
\section{Mapping Kepler orbits into Isochrone orbits}
Newton was stimulated by the fact that both Kepler's orbits under an inverse square force and Hooke's orbits under a linear force law were  ellipses. In the later editions of Principia he showed how the orbits in one could be mapped into orbits in the other. His method is given in Chandrasekhar's book. Here we give a brief characterisation based on the energy equation of an orbit which is rewritten
\be
\2\dot{r}^2+\2h^2/r^2=E+\psi(r);~~~~\psi(r)=GM/r.~~~~\lb{1}
\ee
Newton then takes a new time $\B t(t)$ and a new radius $\B r(r)$ but demands that the angular momentum term remains "the same" i.e. that it becomes $\2h^2/\B r^2$.
\be
\2(d\B r/d\B t)^2+F^2\2h^2/ r^2=F^2\psi(r)+F^2 E;  ~~~~~~F=(d\B r/dr)(dt/d\B t).\lb{2}
\ee
Newton's requirement on the angular momentum term gives $F=r/\B r$.
Notice that the order of the last two terms has been reversed. Newton's method identifies the old and new terms in this switched order. The new energy  is thus $\B E=(r/\Br)^2GM/r$ Since it must be constant we find $r=[\B E/(GM)]\B r^2$. The new potential term $\B\psi$ may now be expressed in terms of the new radius $\B r$ in the form $\B \psi(\B r)=E(GM/\B E)^{-2}~\Br^2$. Thus we have mapped both the angular momentum and the energy equations of Kepler orbits into those of the simple harmonic oscillator.
 
However we do not have to map ALL of the energy term into the potential and vice versa. We can decide to take the new energy as $(1-\la)$ times the old potential term plus $\la$ times the old energy term, provided we take the new potential term to be $(1-\la)$ times the old energy term and $\la$ times the old potential term. On doing this we find
\ba
\B E=(r/\B r)^2[(1-\la)GM/r+\la E],\nn \\
\B\psi(\B r)=(r/\B r)^2[(1-\la)E+\la GM/r]
\ea
We solve the quadratic to find $r(\B r)$ and find that the resultant potential is
\ba
\B \psi(\B r)=G\B M/(b+\B s)+C;~~~\B M=M(2\la-1)\sqrt{E/(\la\B E)};\nn\\
C=\B E(1-\la)/\la:~~~b=\2(1-\la)G M/\sqrt{\la E\B E};~~~\B s=\sqrt{\B r^2+b^2};
\ea
Which is H\'enon's isochrone potential \cite{Hei}. To determine times one must use equation (\ref{2}) to
find the new time $\B t$. We remark that we might have chosen $\la$ to be a function of $r$
but this was not necessary to obtain the isochrone.
% Statistical Mechanics
\section{Statistical Mechanics}
In classical statistical mechanics each quadratic term  in the energy contributes $\2kT$.
For a one dimensional oscillator the kinetic energy averaged over a period $P$ is
\be
\2kT=\int\2 m \dot{x}^2dt/P=\int \2p dx/P=\2J/P=\2J/(dJ/dE)
\ee
Here $J$ is the adiabatic invariant and by inspection it is also the total volume of phase space with energy less than $E$.
Indeed $k \ln [J(<E)]$ is an expression for the entropy of a micro-canonical ensemble recommended by Gibbs. Thus  for such an ensemble the second law of thermodynamics becomes\\
\be
1/(kT) =d \ln [J(<E)]/d E. \lb{3}
\ee
 In an ensemble of simple harmonic oscillators each has $J \propto E$ so the energy $E=kT$ half being contributed by the kinetic energy and half by the potential energy which grows steeply at large $x$. The heat capacity $dE/dT$ is just $k$. \\But now consider a one dimensional H\'enon oscillator with the potential of the isochrone \\$\psi(x)=GM/(b+s);~~s=\sqrt{x^2+b^2}.$ At small $x$ this starts like a simple harmonic oscillator but grows far less steeply at large $x$. For it
 \ba
 J=2GM(-2E)^{-\2}-2\sqrt{GMb};~~~
 \bt=1/(kT)=d\ln J/dE=2GM(-2E)^{-3/2}/J
 \ea
so its heat capacity is
\be
dE/dT=\frac{k}{3\sqrt{-2Eb/(GM)}~-2}
\ee
This rises to infinity as $E$ tends to $-(2/9)GM/b$ from below. Thereafter $C$ is negative and approaches $-\2 k$ at the escape energy $E=0$ see figure 1. There we also draw the interesting
graph of $T(E)$ which increases at first but reaches a maximum and falls to zero at the escape energy.
 \begin{figure}[htbp]
\begin{center}
\includegraphics[width=7cm]{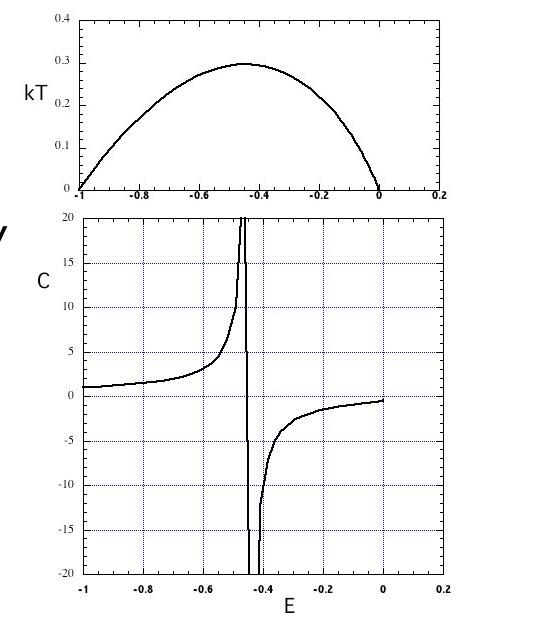}
\caption{ (above)The Temperature of a micro-canonical ensemble of Isochrone oscillators as a function of the energy. The temperature rise is reversed by the large volumes of phase space encountered
as the energy increases towards the escape energy
(below) The heat capacity increases to infinity and then becomes negative where energy increase is associated with a decrease of temperature }
\label{fig1}
\end{center}
\end{figure}
The behaviour of the H\'enon oscillator mimics the behaviour of whole clusters worked on by  
Michel but we now turn to the behaviour of the binary stars he invoked as the central energy source.%Binary Stars
\section{The Temperature of  Binary Stars'  relative motion}
The total energy of a binary star with masses $m_1,m_2;~~m_1+m_2=M$ in a cluster with potential $\psi(r)$ can be written with $E_M$, the energy of its barycentre, and $E_i$ its internal energy. 
\ba
\2 m_1v_1^2+\2 m_2 v_2^2-M\psi-\frac{Gm_1m_2}{r}=\frac{{\bf P}^2}{2M}-M\psi(\B r)+\frac{m_1m_2({\bf v_1-v_2)}^2}{2M}-\frac{Gm_1m_2}{r}\\
E~~~~~~~~~~~~~~~=~~~~~~~E_M~~~~~~+~~~~~~~~~~~~~~~~E_i~~~~~~~~~~~~~~~~~.
\ea
We treat the internal motion of the binary as we treated an oscillator. Its reduced mass is $\mu=m_1m_2/M$; its internal momentum is ${\bf p}=\mu({\bf v_1-v_2})$ and the volume of its internal phase space with energy less than $E_i$ is
\be
J(<E_i)=\int \int 4\pi p^2 dp 4\pi r^2 dr= \3 (\pi G m_1 m_2\sqrt{2\mu})^3)(-E_i)^{-3/2}
\ee
so the internal temperature $T_b$, and heat capacity, $C_i$ of a micro canonical ensemble of such binaries all of the same internal energy is given by \ref{3}
\ba
kT_b=[d\ln J/dE_i]^{-1}(=3/2)^{-1}(-E_i);\nn \\
C_i=dE_i/dT_b=-(3/2)k<0.
\ea
So the temperature is greatest for binaries of large binding energy and all binaries get hotter when cooled and cooler when heated \cite{Pa}.
Now consider binaries interacting with field stars of temperature $T$. Soft binaries with $T_b<T$
will gain heat from such encounters so their binding will decrease and they will get even cooler.
However hard binaries with $T_b>T$ will lose heat get even more tightly bound and thus get hotter.
It was to such interactions as well as binary formation that H\'enon appealed in his theory of star cluster evolution. This thermodynamics gives a pretty explanation of Heggie's \cite{Heg} laws of binary interaction.
Spitzer \cite{Sp} emphasised the importance of mass segregation in clusters. Stars of greater mass in local equipartition of kinetic energy move more slowly, thus they sink towards the centre and gain kinetic energy thereby. Near the centre they may well form binaries and in smaller clusters the binding energy of a single massive binary may contribute a large fraction of the total binding energy of the cluster.
van Albada \cite{vA} found this in his early simulations of small clusters and Sverre Aarseth \cite{Aa} found that single 
binaries could dominate even quite large clusters.
%Newton's N-body problem and the Isochrone
\section{Newton's N-body problem and the Isochrone}
Chemists know about mass-weighted coordinates. These allow one to treat motion of bodies of different mass subject to simple harmonic interactions  with the same ease that one treats bodies of equal mass. However I shall save you that complication and consider Newton's problem but with all $N$ bodies of equal mass, $m=M/N$. The potential energy of all the binary harmonic interactions is given by
\be
V=-G' m^2 \sum {\sum{({\bf r_i-r_j})^2}}/4=-G'mM\sum{({\bf r_i-\B r})^2}/2; ~~~~{\bf \B r}=\sum{{\bf r_i}}/M
\ee
and the kinetic energy, $K$, is
\be
K=\2 m \sum( {\bf \dot{r_i}^2})
\ee
Now use the barycentre, ${\bf \B r}$,  as origin and work in $3N$ dimensions with\\ ${\bf  R=(r_1,r_2,r_3....,r_N)}$, so ${\bf R}$ gives the positions of all $N$ bodies and ${\bf \dot{R}}$
gives all their velocities. Notice that $V=-\2G'mM {\bf R}^2$ and $K=\2m{\bf\dot{R}^2}$. The initial 
positions define an initial ${\bf R}$ and the initial velocities define the initial ${\bf \dot{R}}$ and those two Vectors define a two dimensional plane in the $3N$ dimensional space. However the force is central in the $3N$ dimensional sense so the motion never leaves the initial plane. This beautiful 
property is still true if $V$ is any function of $R=\sqrt{{\bf R}^2}$. In particular one can take $V=-\2GM^2/R$ or make $V=-G'M^2
/(Nb+\sqrt{R^2+N^2b^2})$ the Isochrone potential as was considered in \cite{LB^2}.
The orbit in the plane of the motion is Michel H\'enon's orbit but to find the orbit of any one body one
must project out the $3N-3$ unwanted coordinates. One gets the projection of an isochrone's rosette
i.e. a rosette between two similar ellipses.

Michel took great delight in mathematical beauty. I  hope that some of these extensions of his model
still transmit that delight.

\end{document}